\documentclass[12pt]{article}
\usepackage{amssymb}

\usepackage{amsmath}
\usepackage{graphicx}

\begin{document}

\author{I. Agnolin$^{*}$, N.P. Kruyt$^\dag$}
\title{On the elastic moduli of two-dimensional assemblies of disks: relevance and
modeling of fluctuations in particle displacements and rotations}
\date{}
\maketitle
\begin{center}
$^*$Groupe Mati\`{e}re Condens\'{e}e et Mat\'{e}riaux, University of Rennes I%
\\[0pt]
Bat.11A, Campus Beaulieu, 35042 Rennes, France\\
$^\dag$Department of Mechanical Engineering, University of Twente\\[0pt]
P.O. Box 217, 7500~AE Enschede, The Netherlands\\[0pt]
n.p.kruyt@utwente.nl, fax +31(0)534893695
\end{center}

\begin{abstract}
% Text of abstract
We determine the elastic moduli of two-dimensional assemblies of
disks by computer simulations. The disks interact through elastic
contact forces, that oppose the relative displacement at the contact
points by means of a normal and a tangential stiffness, both taken
constant. Our simulations confirm that the uniform strain assumption
results in inaccurate predictions of the elastic moduli, since large
fluctuations in particle displacements and rotations occur. We
phrase their contribution in terms of the relative displacement they
induce at the contact points. We show that the fluctuations that
determine the equivalent continuum behavior depend on the average
geometry of the assembly. We further separate the contributions from
the center displacement and the particle rotation. The fluctuations
result in a relaxation of the system, but along the tangential
direction the relaxation is generally entirely due to rotations. We
consider two theoretical formulations for predicting the elastic
moduli that include the fluctuations, namely the
``pair-fluctuation'' and the ``particle-fluctuation'' method. They
are both based on the equilibrium of a small subassembly, which is
considered representative of the average structure. We investigate
the corresponding predictions of the elastic moduli over a range of
coordination numbers and of ratios between tangential and normal
stiffness. We find a significant improvement with respect to the
uniform strain theory. Furthermore, the dependence of the
fluctuations on coordination number and ratio of tangential to
normal stiffness is qualitatively captured.
\end{abstract}



%\author{The Author}
%\title{The Title }
%\date{The Date }
%\maketitle
%\tableofcontents



\section{Introduction}

This study investigates the elastic mechanical properties of dense
random isotropic two-dimensional assemblies of disks. Our work is
framed in the context of micromechanics, which focuses on the
relation between macroscopic behavior and microscopic interactions.
At the macroscopic scale, stress and strain are measured, whose
relation is determined by the elastic moduli of the equivalent
continuum. At the microscale, forces arise between contacting
particles that for quasi-static deformations must satisfy the
balance of force and moment for all grains. \newline \indent In the
context of elasticity, contact forces oppose the relative
displacement between contacting grains by means of a contact
stiffness. The work of Poritzky \cite{Poritsky} about contacting
thin disks reveals for its normal component a dependence on the
overlap that rapidly weakens with the confining pressure. We assume
for our assemblies the limit case of constant normal stiffness.
Since we focus on small displacements, we also consider the
tangential stiffness constant \cite {Mindlin2,Johnson}. By ignoring
all inhomogeneity in the contact stiffness, we specifically focus on
the effects of geometric disorder.

Given the geometry of the contact network, the constitutive law for
the contact forces determines the macroscopic mechanical properties
once the relative displacements at the contact points are known.
Many micromechanical studies \cite{Rothenburg1980, Digby,
Christoffersen, Walton, BathurstRothenburg1988a,
BathurstRothenburg1988b, Changetal1990, Changetal1995, Cambouetal}
use the so-called `average strain hypothesis', where the relative
displacements reduce to the contribution from the average strain.
The resulting prediction of the elastic moduli is inaccurate
\cite{Makseetal,RothenburgKruyt2001,Suiker}, particularly for the
shear modulus, as in general grains undergo additional displacements
in order to attain equilibrium. More sophisticated predictions have
recently been developed that incorporate such fluctuations, in
\cite{Jks05} for three-dimensional systems and in
\cite{KruytRothenburg2004} and \cite{Agnolinetal2005} for
two-dimensional ones. They are based on the idea that even though
the deformation is not uniform, its fluctuations are strongly
correlated. The study by Koenders \cite{Koenders} suggests that they
occur with correlation lengths in the order of a few diameters,
which implies that subassemblies of such a size already contain the
essential features of the global structure.

The validity of the theoretical predictions in
\cite{KruytRothenburg2004} and \cite{Agnolinetal2005} is
investigated here by comparison with the elastic moduli computed by
means of Discrete Element Method (\emph{DEM}) simulations. The
analysis is performed for various ratios of tangential to normal
stiffness and coordination numbers. We consider assemblies with both
larger and smaller coordination number than the onset of
iso-staticity for disordered frictionless systems. This onset equals
$4$ in two dimensions and is recurrent in numerical simulations, as
dense assemblies are usually obtained by means of an initial
frictionless compression. In experiments \cite{Bideau} on hard
disks, coordination numbers smaller and larger than $4$ are due, in
turn, to the presence of friction and of ordered structures. In DEM
simulations, disordered assemblies with larger coordination numbers
than the onset of isostacity can be obtained, for example by
neglecting in the constitutive law for the contact force the
increase in the normal stiffness with the interpenetration or by
applying a large pressure. Such systems are mainly of academic
interest, particularly in the development of statistical approaches
aimed at predicting the evolution of disordered systems. Also in
three dimensions the scientific literature concentrates on samples
with larger coordination numbers than the frictionless onset of
iso-staticity, which equals $6$ in that case. However, studies
concerned with the issue of numerically reproducing experimental
results \cite{CundallJenkins}, \cite{AgnPrep} emphasize that lower
ones might be relevant to practical purposes.

The outline of the study is as follows. Firstly, the basic
micromechanical quantities of interest are defined in Section
\ref{Micromechanics}. Then, a concise description is given in
Section \ref{Theories} of theoretical approaches for predicting the
elastic moduli based, in turn, on the average strain assumption and
on the inclusion of displacement fluctuations. The performed DEM
simulations are described in Section \ref{DEM simulations}. The
corresponding results are analyzed in Section \ref{Analysis} and
compared to the results the theoretical predictions. The final
section is dedicated to discussion of the results.

\section{Micromechanics}

\label{Micromechanics}
We consider the contact between disks $p$ and $q$ of radius $R^{p}$ and $%
R^{q}$, respectively. The contact is identified by the unit vector $n_{i}^{pq}$ that points
outwards from $p$ along the line that joins the centers. The unit vector $%
t_{i}^{pq}$ is tangent to the contact (see Figure \ref{Contact
geometry}). In components,
\begin{eqnarray*}
\mathbf{n}^{pq} &=&(\cos \theta ^{pq},\sin \theta ^{pq}), \\
\mathbf{t}^{pq} &=&(-\sin \theta ^{pq},\cos \theta ^{pq}),
\end{eqnarray*}
where $\theta ^{pq}$ is the \emph{contact orientation}, counted
counterclockwise from the horizontal axis. For future reference, we
also define the \emph{branch vector} $l_{i}^{pq}$, that joins the
centre of particle $p$ to that of particle $q$ pointing outwards
from $p$, i.e.
\begin{equation*}
l_{i}^{pq}=\left( R^{p}+R^{q}\right) n_{i}^{pq}.
\end{equation*}
As in monodisperse two-dimensional assemblies crystallization occurs, a log-normal distribution for the particle radii is adopted.
\begin{figure}[h]
\centering
\includegraphics[width=6cm]{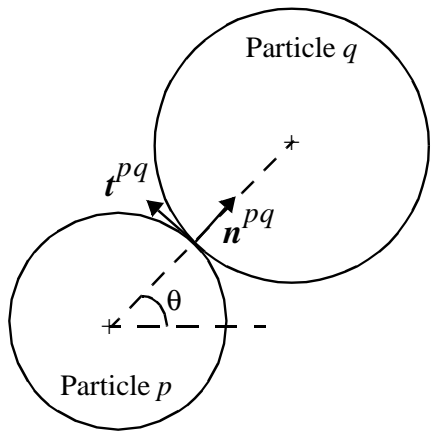}
\caption{Contact geometry: contact orientation and normal and
tangential vectors to the contact.}
\label{Contact geometry}
\end{figure}
Contacting particles interact by means of contact forces. We denote by $%
f_{i}^{pq}$ the $i$-th component of the force exerted on particle
$p$ by particle $q$. It has normal and tangential component to the
contact $f_{n}^{pq}$ and $f_{t}^{pq}$, i.e.
\begin{eqnarray*}
f_{n}^{pq}&=&f_{i}^{pq}n_{i}^{pq},\\
f_{t}^{pq}&=&f_{i}^{pq}t_{i}^{pq}.
\end{eqnarray*}
In the hypothesis of quasi-static deformations, contact forces satisfy the balance of force and moment on each grain. For instance, on
particle $p$,
\begin{eqnarray*}
\sum_{q}f_{i}^{pq}&=&0,  \notag \\
\sum_{q}e{_{jk}}R^{p}n{_{j}^{pq}}f{_{k}^{pq}}&=&0,  \label{Equilibrium equations}
\end{eqnarray*}
where the sum is over particles $q$ that are in contact with it and
$ e_{jk}$ is the two-dimensional permutation tensor. Contact forces
oppose the relative displacement between contacting particles by
means of a contact
stiffness, whose normal and tangential component we denote, in turn, by $%
k_{n}$ and $k_{t}$, both considered constant. If we denote by $\Delta
_{i}^{pq}$ the relative displacement between particles $p$ and $q$ that are
in contact, and by $\Delta _{n}^{pq}$ and $\Delta _{t}^{pq}$ its normal and
tangential component, we have
\begin{eqnarray}
f_{n}^{pq}&=&k_{n}\Delta _{n}^{pq},  \notag \\
f_{t}^{pq}&=&k_{t}\Delta _{t}^{pq},  \notag \\
f_{i}^{pq}&=&f_{n}^{pq}n_{i}^{pq}+f_{t}^{pq}t_{i}^{pq}.
\label{Contactconstitutiverelation}
\end{eqnarray}
The theory of contact elasticity predicts the decay with the distance from the contact zone of the effects of contact interactions. For small enough deformations and stiff enough particles, as we assume to be the case here, contact interactions can be assumed to be confined to a contact point, and the grains kinematics to be approximated by that of rigid bodies. Therefore,
\begin{equation*}
\Delta _{i}^{pq}=U_{i}^{q}-U_{i}^{p}+e_{ij}\left( R^{q}\Omega
^{q}+R^{p}\Omega ^{p}\right) n_{i}^{pq},  \label{Delta}
\end{equation*}
where $U_{i}^{p}$ and $U_{i}^{p}$ are the displacement of the centre of
particle $p$ and $q$, $\Omega ^{p}$ and $\Omega ^{q}$ their rotation.

At the macroscopic, continuum level, the relevant quantities are the stress tensor $\sigma_{ij}$ and the strain
tensor $\epsilon_{ij}$, whose components are taken positive in compression.
Contact forces and the geometry of the particle arrangement determine the
expression of the former, as \cite{Drescher, Strack, RothenburgSelvadurai}
\begin{equation}
\sigma _{ij}=\frac{1}{S}\sum_{\theta _{g}}\sum_{c\in C(\theta
_{g})}f_{i}^{c}l_{j}^{c},  \label{StressTheta}
\end{equation}
that is the average over the area of interest $S$ of Cauchy's stress
\cite {Love}. The orientation $\theta _{g}$ varies between $0$ and
$\pi $. Finally, the superscript $pq$ referring to contacting
particles has been replaced by for corresponding contacts $c$.
Expression (\ref {StressTheta}) emphasizes the dependence of the
macroscopic behavior on the average force over equally oriented
contacts. Experimental observations and numerical simulations \cite
{BathurstRothenburg1990,Calvetti,RothenburgBathurst1989} suggest for
it the same dependence on the contact orientation as for the effects
of the average strain. That is, compressive forces are the larger
the closer the contact orientation is to the direction of major
compression, and tangential forces have their maximum at contacts
oriented at 45 degrees from it.

Primary geometrical characteristics of granular assemblies are coordination
number $\Gamma $, i.e. the average number of contacts per particle, contact density $%
n_{S}$, i.e. the average number of particles per unit surface, and the
contact distribution function $E(\theta )$ \cite{Horne}, defined such that $%
E(\theta )d\theta $ gives the probability of finding a contact with
orientation $\theta$ in the interval $(\theta ,\theta +d\theta )$,
$\theta \in (0,\pi )$. In the case of \emph{isotropic} assemblies,
as considered here,
the contact distribution function becomes $E(\theta )=1/\pi $%
. With the use of such quantities and by denoting averages over
equally oriented contacts by overbars, expression
(\ref{StressTheta}) transforms into the integral
\begin{equation}
\sigma _{ij}=\frac{n_{S}\Gamma}{2} \int_{0}^{\pi }E(\theta )\overline{%
f_{i}l_{j}}(\theta )d\theta .
\label{Expression for stress tensor (continuous form}
\end{equation}
The elastic stiffness tensor relates stress and strain. In isotropic
systems, it is fully described by the effective bulk modulus $K$ and shear
modulus $G$, as
\begin{eqnarray}
\sigma _{11}+\sigma _{22}=2K(\epsilon _{11}+\epsilon _{22})  \notag \\
\sigma_{11}-\sigma _{22}=2G(\epsilon _{11}-\epsilon _{22}).
\label{Definition moduli}
\end{eqnarray}
We determine them by means of DEM computer simulations, by applying,
in order, an isotropic compressive deformation $\epsilon _{ij}^{K}$
and a shear deformation $\epsilon _{ij}^{G}$:
\begin{equation}
\epsilon _{ij}^{K}=\epsilon _{0}\left(
\begin{array}{cc}
1 & 0 \\
0 & 1
\end{array}
\right), \qquad \epsilon _{ij}^{G}=\epsilon _{0}\left(
\begin{array}{cc}
1 & 0 \\
0 & -1
\end{array}
\right) ,  \label{Loading paths}
\end{equation}
and measuring the corresponding stress response. By $\epsilon _{0}$
we denote a magnitude of the imposed strain.

\section{Theoretical modeling}

\label{Theories}
\begin{figure}[h]
\centering
\includegraphics[width=12cm]{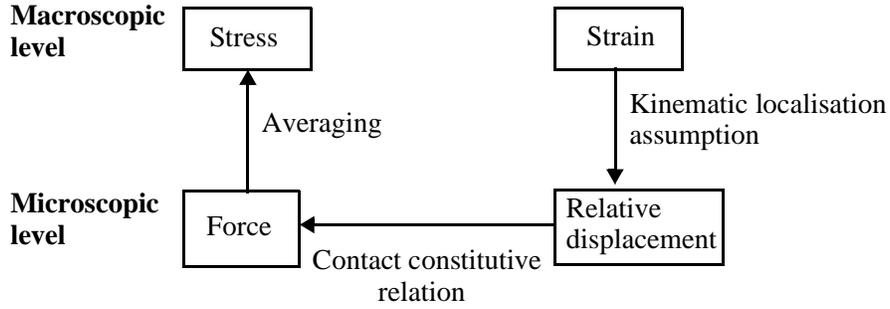}
\caption{Kinematic localisation assumption.}
\label{Localisation}
\end{figure}

If the deformation $\epsilon _{ij}$ is prescribed, the theoretical
prediction of the elastic moduli requires that of the corresponding
stress tensor $\sigma _{ij}$. Due to (\ref
{Contactconstitutiverelation}), this in turn requires a kinematic
localization assumption, that is, the expression of the relative
displacements between contacting particles as a function of
$\epsilon _{ij}$. This procedure is depicted in Figure
\ref{Localisation}. Various kinematic localization assumptions are
considered in this section, such as the uniform strain assumption
and more sophisticated approaches that account for fluctuations.
\subsection{Uniform strain}

\label{Uniform strain} In the hypothesis of uniform strain the
relative displacement between contacting particles becomes
\begin{equation*}
\Delta _{i}^{pq}=\epsilon _{ij}l_{j}^{pq}.
\end{equation*}
Its average over equally oriented contacts in the case of isotropic
compression is isotropic. Only its normal component $\Delta
_{n}^{\epsilon ,is}$ differs from zero, namely
\begin{equation}
\Delta _{n}^{\epsilon}=\epsilon _{0}\overline{l}, \label{Deltais}
\end{equation}
where $\overline{l}$ is the average length of the branch vector. In
the case of shear, the average normal and tangential component
$\Delta _{n}^{\epsilon }$ and $\Delta _{t}^{\epsilon }$ of $\Delta
_{i}^{pq}$ over equally oriented contacts are, in turn,
\begin{eqnarray}
\Delta _{n}^{\epsilon }(\theta ) &=&\epsilon _{0}\overline{l}\cos 2\theta ,  \notag \\
\Delta _{t}^{\epsilon }(\theta ) &=&-\epsilon _{0}\overline{l}\sin
2\theta. \label{DeltaE}
\end{eqnarray}
For isotropic assemblies of disks, the average strain assumption results in the bulk and shear moduli $%
K^{\epsilon }$ and $G^{\epsilon }$ \cite{BathurstRothenburg1988a}:
\begin{eqnarray}
\frac{K^{\epsilon }}{k_{n}} &=&\frac{n_{S}\Gamma }{8}\;\overline{l^{2}}
\notag \\
\frac{G^{\epsilon }}{k_{n}} &=&\frac{n_{S}\Gamma }{16}\left[ 1+\frac{k_{t}}{%
k_{n}}\right] \overline{\text{ }l^{2}},
\label{Moduliaccordingtouniform strain}
\end{eqnarray}
that are upper bounds. The symbol $\overline{l^{2}}$ denotes the
average over all contacts of the squared length of the branch
vector. As discussed in \cite{KruytRothenburg2001}, polydispersity
makes the average length of the branch vector differ in general from
the average diameter.  For our size distribution $\overline{l}$ is
about $3\%$ larger than the average diameter, and $\overline{l^{2}}$
is about $15\%$ larger than the average diameter squared.

\subsection{Approaches that incorporate displacement fluctuations}

\label{Fluctuation methods}

Two approaches are considered that incorporate the fluctuations into
the prediction of the elastic moduli, namely the
particle-fluctuation (1PF) and the pair-fluctuation method (PF).
They are discussed in detail in \cite {KruytRothenburg2004} and in
\cite{Agnolinetal2005}, respectively. In \cite
{KruytRothenburg2004}, the 1PF approach is applied numerically and
proven to give upper bounds to the effective moduli, while in
\cite{Agnolinetal2005} the analytical solution that corresponds to
the PF method is presented. Both approaches deal with the issue of
equilibrium of a small assembly, made of a chosen particle $A$ and
of a contacting pair $AB$, respectively, surrounded by their first
neighbors. Therefore, in these models fluctuations are assumed to
occur with correlation lengths in the order of 3 or 4 diameters.

In the more general case, the relative displacement between
contacting particles can be decomposed into the contribution from
averages and fluctuations, that is, from the imposed strain and the
average particle rotation, on one side, and from fluctuations in
particle displacements and rotations, on the other. In the
absence of macroscopic rotation, as it is the case here, in an
isotropic system the
average particle rotation is zero \cite{Jenkins99}. If we denote by $\tilde{u}_{i}^{p}$ and $%
\tilde{u}_{i}^{q}$ the fluctuation of contacting particles $p$ and
$q$ in the center displacement, and by $\tilde{\omega}^{p}$ and
$\tilde{\omega}^{q}$ the fluctuations in rotations, we can write:
\begin{equation*}
\Delta _{i}^{pq}=\epsilon _{ij}l_{j}^{pq}+\left(\tilde{u}_{i}^{q}-\tilde{u}%
_{i}^{p}\right)-\left( R^{q}\tilde{\omega}^{q}+R^{p}\tilde{\omega}%
^{p}\right) t_{i}^{pq}.
\end{equation*}

\begin{figure}[h]
\centering
\includegraphics[width=6cm]{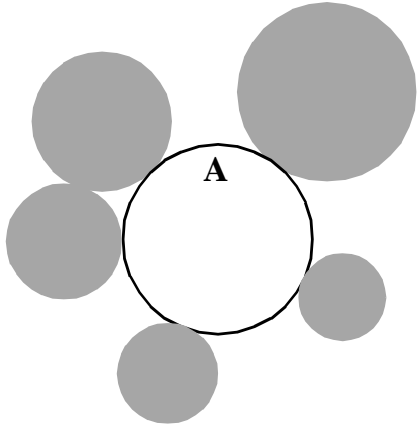}
\caption{Small assembly considered in the "particle-fluctuation" method,
centered on particle $A$. The deformation of the particles shown in gray is
according to uniform strain, while $A$ is allowed to fluctuate. Its
fluctuations stem from the balance of force and moment on it.}
\label{PFmethod}
\end{figure}

In the 1PF and the PF approach, only the chosen
particle or pair is allowed to fluctuate, while their neighborhood
is compelled to move according to the average strain. This is
depicted in Figure \ref{PFmethod} for the 1PF case. As a result, in
the 1PF method the relative displacement between particle $A$ and
its $r$-th neighbor reduces to
\begin{equation*}
\Delta _{i}^{Ar}=\epsilon _{ij}l_{j}^{Ar}-\tilde{u}_{i}^{A}-R^{A}\tilde{%
\omega}^{A} t_{i}^{Ar},
\end{equation*}
so that the three equilibrium equations for particle $A$ can be solved for
its three unknown fluctuations. In the pair-fluctuation method,
\begin{eqnarray*}
\Delta _{i}^{AB} &=&\epsilon _{ij}l_{j}^{AB}+\left(\tilde{u}_{i}^{B}-\tilde{u%
}_{i}^{A}\right)-\left( R^{A}\tilde{\omega}^{A}+R^{B}\tilde{\omega}%
^{B}\right) t_{i}^{AB},
\end{eqnarray*}
but
\begin{eqnarray*}
\Delta _{i}^{Ar} &=&\epsilon _{ij}l_{j}^{Ar}-\tilde{u}_{i}^{A}-R^{A}\tilde{%
\omega}^{A} t_{i}^{Ar},  \notag \\
\Delta _{i}^{Bs} &=&\epsilon _{ij}l_{j}^{Bs}-\tilde{u}_{i}^{B}-R^{B}\tilde{%
\omega}^{B} t_{i}^{Bs},
\end{eqnarray*}
where $r$ denotes a neighbor of particle $A$ different from $B$, and $s$ a
neighbor of particle $B$ different from $A$. The six equilibrium equations
of particles $A$ and $B$ can be solved for the corresponding six
fluctuations. In both cases, the small size of the solving system permits an
analytical formulation.

In this work, the equilibrium equations are solved numerically,
particle by particle and pair by pair for the 1PF and the PF
approach, respectively, for the corresponding fluctuations. Using
the procedure sketched in Figure \ref{Localisation}, the
corresponding contact forces are computed and the stress and
elastic moduli estimated. The resulting fluctuations are also
analyzed in terms of the associated deformation mechanisms in the
sections that follow.

\section{DEM simulations}

\label{DEM simulations}

DEM simulations have been performed of large isotropic assemblies of
50,000 disks, with radii from a lognormal distribution. Periodic
boundaries have been employed to reduce boundary effects.

In the DEM method, the deformation of the assemblies is computed by
numerically integrating in time the equations of motion of all
particles, and the results can be employed to analyze the actual
deformation mechanisms.

Initial equilibrium states with coordination numbers $\Gamma=3.5$,
$\Gamma=4$ and $\Gamma=5$ have been first prepared. To obtain the
first one, a succession of frictionless and frictional compressions
has been employed. The second one results from the isotropic
compression of a frictionless gas, and the third one results from an
additional isotropic compression.

The assemblies obtained this way have been subjected to the two
strain paths specified by eqn.(\ref{Loading paths}). Then, the
corresponding stress response has been computed in order to
determine the effective bulk and shear moduli $K$ and $G$. Such
simulations have been performed with bonded contacts: that is,
neither contact creation nor disruption has been considered, which
is appropriate for studying elastic
behavior at small strains. At each coordination number, different ratios $%
k_{t}/k_{n}$ have been used, in the range between 0.05 and 1.0.

\section{Micromechanical analysis}

\label{Analysis}

In this section, the results from the DEM simulations and the
theoretical approaches introduced in Section \ref{Theories} are
analyzed. The comparison between the elastic moduli from the DEM
simulations and the uniform strain assumption emphasizes the
relevance of displacement fluctuations. The quality of the estimates
of the elastic moduli given by the 1PF and the PF methods is also
evaluated. Then, the deformation mechanisms induced by the
fluctuations at the macroscopic scale are analyzed, and the
performance of the 1PF and PF approaches interpreted in terms of
their capability of capturing them.

\subsection{Moduli}
\begin{figure}[h]
\centering
\includegraphics[width=13cm]{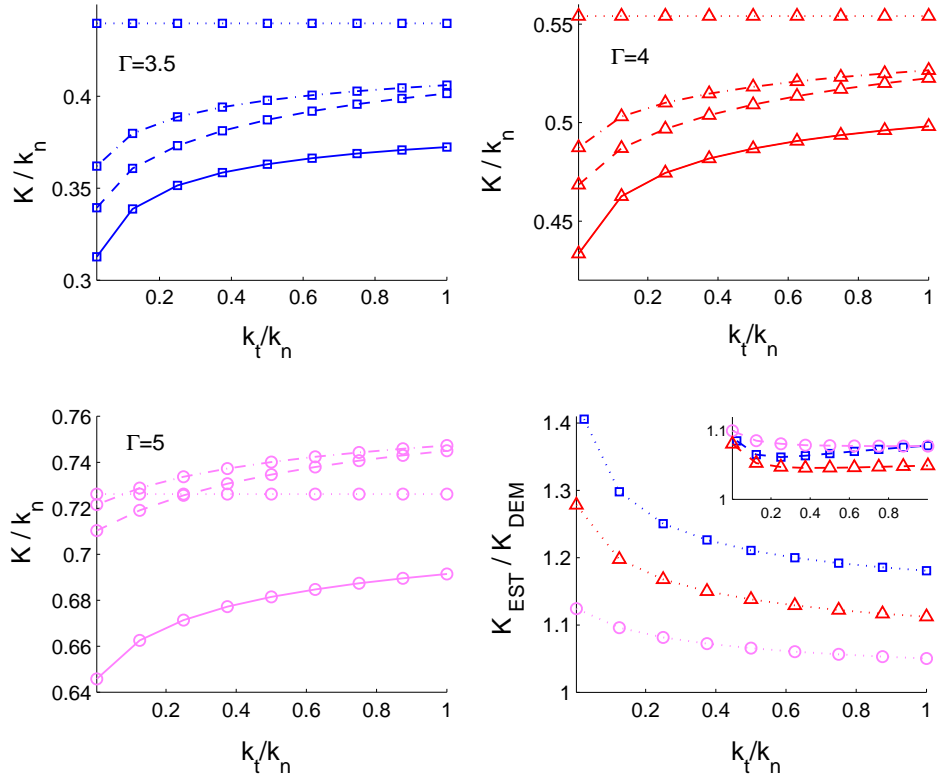}
\caption{Bulk modulus $K$: from DEM simulations (solid line), uniform
strain assumption (dotted line), 1PF (dashed) and PF (dashed-dotted)
approaches. Results for three coordination numbers $\Gamma $: $\Gamma
=3.5\,(\Box )$, $\Gamma =4$\thinspace ($\triangle $), $\Gamma =5\,(\bigcirc
) $ and various stiffness ratios $k_{t}/k_{n}$. Bottom right: estimate performance for uniform strain assumption (dotted) and 1PF approach
(dashed), for $\Gamma =3.5 ,4.0 , 5.0$.}
\label{Bulk}
\end{figure}
\begin{figure}[h]
\centering
\includegraphics[width=12cm]{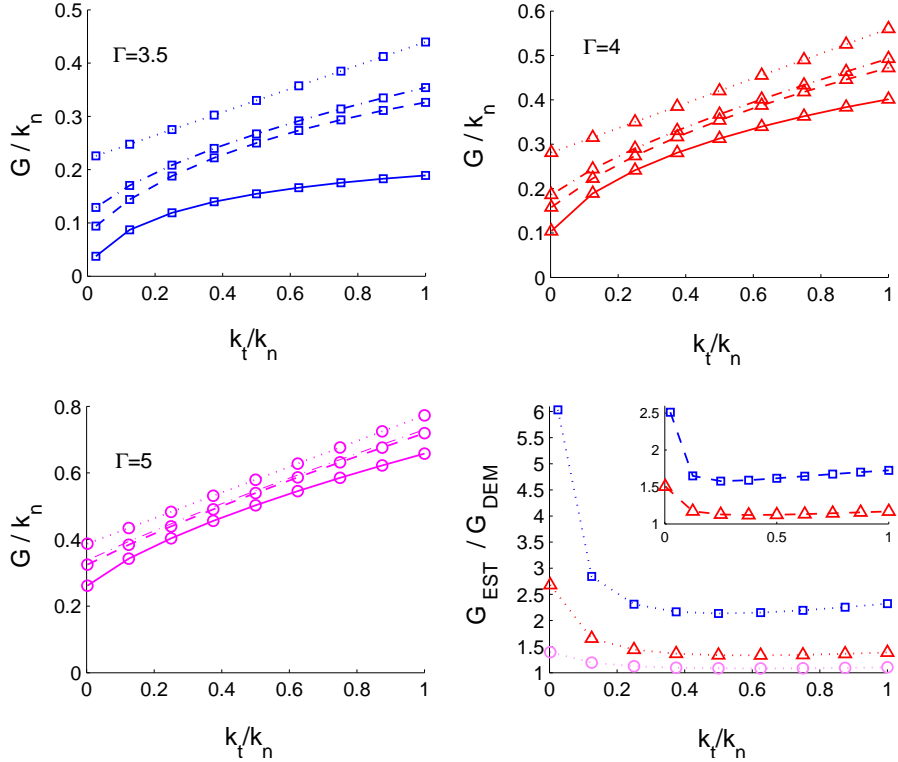}
\caption{Shear modulus G. Same symbols as in Figure \ref{Bulk}.}
\label{Shear}
\end{figure}

The bulk and shear moduli that result from the DEM simulations and the
theoretical approaches discussed in Section \ref{Theories} are compared in
Figures \ref{Bulk} and \ref{Shear}. They are presented in the dimensionless
form $K/k_{n}$ and $G/k_{n}$, as function of the ratio $k_{t}/k_{n}$ and for
the three chosen coordination numbers. In the fourth frame of the two
figures, the quality of the estimates is plotted in terms of their ratio to
the effective moduli.
The average strain prediction performs poorly, especially at small coordination numbers. The inclusion of the
fluctuations with as few degrees of freedom as in the 1PF and PF methods
does not guarantee an improvement with respect to the average strain
assumption, as the case of the bulk modulus at $\Gamma =5$ proves.

The DEM simulations point out the dependence of the bulk modulus on
the tangential stiffness, ignored by the average strain assumption,
but captured once fluctuations are accounted for. The reason tofor
such a dependence lies in the fact that even though the tangential
forces are zero on average in hydrostatic compression, they are
required at the particle level in order for the forces and moments
to balance. By opposing the relative displacements along the
tangential direction, the tangential forces stiffen the assembly,
the more so the larger the tangential stiffness. The general belief
that the average strain assumption gives a good approximation of the
bulk modulus seems hardly acceptable, at least below the onset of
iso-staticity for frictionless systems and at small ratios
$k_{t}/k_{n}$. Regarding the shear modulus, the inclusion of the
fluctuations into the prediction results in a significant
improvement with respect to the average strain assumption, although
important deviations from the DEM simulations remain, especially for
low coordination number and low stiffness ratios $k_{t}/k_{n}$.

\subsection{Relative displacements}

\label{Relative displacements}

At the macroscopic level, only the average over equally oriented
contacts of the relative displacement between contacting particles
is important, as follows from (\ref{Contactconstitutiverelation})
and (\ref {StressTheta}). We label by $\tilde{\Delta}_{i}^{pq}$ the
relative displacement between contacting grains $p$ and $q$ due to
their fluctuations:
\begin{equation*}
\tilde{\Delta}_{i}^{pq}\doteq \tilde{u}_{i}^{q}-\tilde{u}_{i}^{p}-\left(
R^{p}\tilde{\omega}^{p}+R^{q}\tilde{\omega}^{q}\right) t_{i}^{pq},
\end{equation*}
and by $\tilde{\Delta}_{n}^{pq}$ and $\tilde{\Delta}_{t}^{pq}$ its normal
and tangential component,
\begin{eqnarray*}
\widetilde{\Delta }_{n}^{pq} &=&\left( \widetilde{u}_{j}^{q}-\widetilde{u}%
_{j}^{p}\right) n_{j}^{pq} \\
\widetilde{\Delta }_{t}^{pq} &=&\left( \widetilde{u}_{j}^{q}-\widetilde{u}%
_{j}^{p}\right) t_{j}^{pq}-\left( R^{q}\widetilde{\omega }^{p}+R^{p}%
\widetilde{\omega }^{q}\right).
\end{eqnarray*}
\begin{figure}[h]
\centering
\includegraphics[width=10cm]{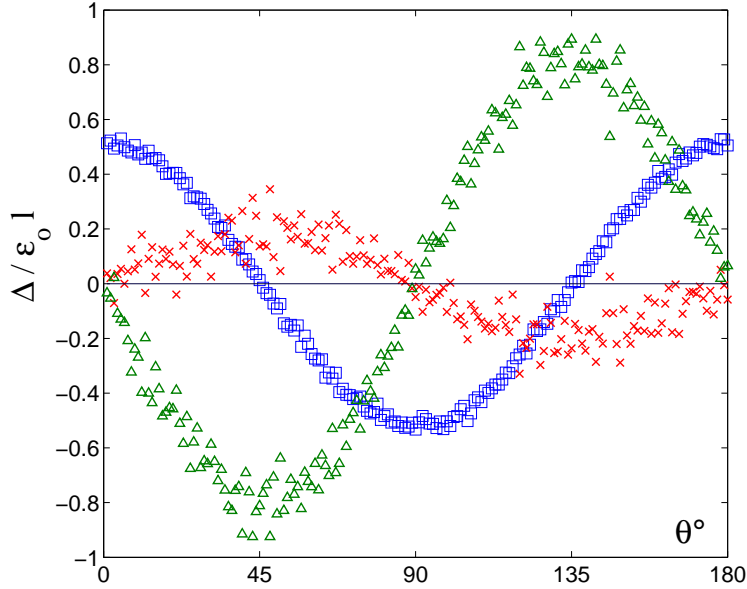}
\caption{Relative displacements between contacting particles averaged over
equally oriented contacts: normal component ($\Box $), tangential component
due to center displacements ($\triangle $) and rotations ($\text{x}$).
Results for shear deformation at coordination number $\Gamma =3.5$ and
stiffness ratio $k_{t}/k_{n}=0.5$.}
\label{Displ35}
\end{figure}
When an isotropic compression is applied, only the former contributes to the
stress. Its distribution is uniform and has average $\overline{\tilde{\Delta%
}}_{n}$, such that
\begin{equation}
\overline{\tilde{\Delta}}_{n}=\beta _{n}\Delta _{n}^{\epsilon},
\label{alphais}
\end{equation}
where $\Delta _{n}^{\epsilon}$ is given by expression
(\ref{Deltais}). As the fluctuations relax the system, $\beta_{n}$
is negative. It follows after some algebra from (\ref{StressTheta}),
the first of (\ref{Moduliaccordingtouniform strain}) and
(\ref{alphais}) that
\begin{equation}
K=\left(1+\beta_{n}\right)K^{\epsilon}.
\end{equation}
Corresponding values of $K^{\epsilon}/K$ are shown in
the fourth frame of Figure (\ref{Bulk}).

Typical group averages for the case of shear loading are shown in
Figure \ref{Displ35}, where along the tangential direction the
contributions from the center displacement and the rotation have
been separated. The average normal and tangential relative
displacements induced by the fluctuations over equally oriented
contacts are proportional to those of eqn.(\ref{DeltaE}), induced by
the average strain and aligned with them. Hence we can write \cite
{KruytRothenburg2001}:
\begin{eqnarray*}
\overline{\tilde{\Delta}}_{n}(\theta ) &=&\alpha _{n}\Delta
_{n}^{\epsilon }(\theta ) \\
\overline{\tilde{\Delta}}_{t}(\theta ) &=&\alpha _{t}\Delta _{t}^{\epsilon
}(\theta ).
\end{eqnarray*}
The contributions from the particle displacements and rotations to $%
\overline{\tilde{\Delta}}_{t}$ are characterized by analogous coefficients $%
\alpha _{t}^{u}$ and $\alpha _{t}^{\omega }$, such that
\begin{equation*}
\alpha _{t}=\alpha _{t}^{u}+\alpha _{t}^{\omega }.
\end{equation*}
Considerations analogous to those arisen from expression (\ref{prodFluct}) allow one to write the effective shear modulus as
\begin{equation*}
\frac{G}{k_{n}}=(1+\alpha _{n})\left[ 1+\frac{(1+\alpha _{t})k_{t}}{%
(1+\alpha _{n})k_{n}}\right] \frac{n_{S}\Gamma }{16}\text{ }\overline{%
l^{2}}.
\end{equation*}

\subsection{Magnitude of observed and predicted fluctuations}
\begin{figure}[h]
\centering
\includegraphics[width=13cm]{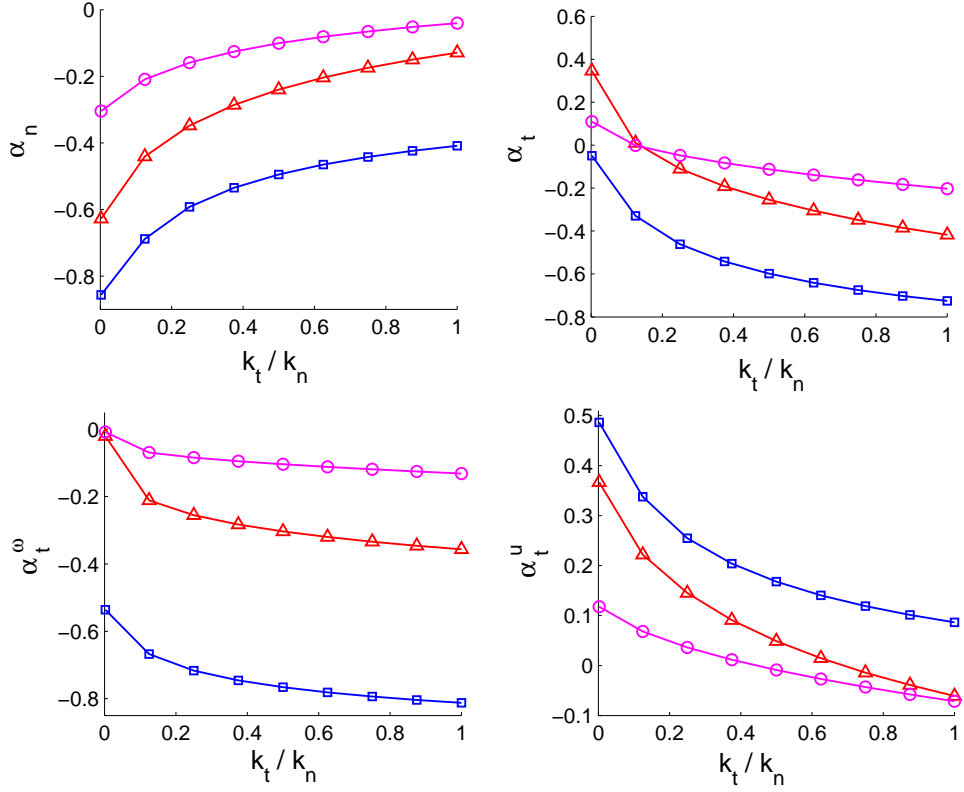}
\caption{Magnitude of fluctuations in the relative displacements in terms
of $\alpha_n^{sh}$, $\alpha_t$, $\alpha_t^\omega$ and $\alpha_t^u$. Results
from DEM simulations for three coordination numbers $\Gamma$: $%
\Gamma=3.5\,(\Box)$, $\Gamma=4$\,($\triangle$), $\Gamma=5\, (\bigcirc)$ and
various stiffness ratios $k_t/k_n$.}
\label{alphaDEM}
\end{figure}
In the case of isotropic compression, Figure \ref{Bulk} shows that
the prediction of $\beta_{n}$ is satisfactory.

As regards shear, the dependence of $\alpha _{n}$, $\alpha _{t}$, $\alpha _{t}^{u}$ and $%
\alpha _{t}^{\omega }$ on coordination number and stiffness ratio $%
k_{t}/k_{n}$ is shown in Figure \ref{alphaDEM}. Negative numerical
factors mean a relaxation with respect to the average strain
assumption. The magnitude of the fluctuations decreases with
increasing coordination number and, as expected, is  more important
when the system is undergoing shear deformation. Both the normal and
the tangential component of the fluctuations relax the system with
respect to the average strain assumption, but along the tangential
direction the relaxation is almost exclusively due to the particle
rotation. On the contrary, the center displacements generally induce
average relative displacements of the same sign as those due to the
average strain.
\begin{figure}[h]
\centering
\includegraphics[width=14cm]{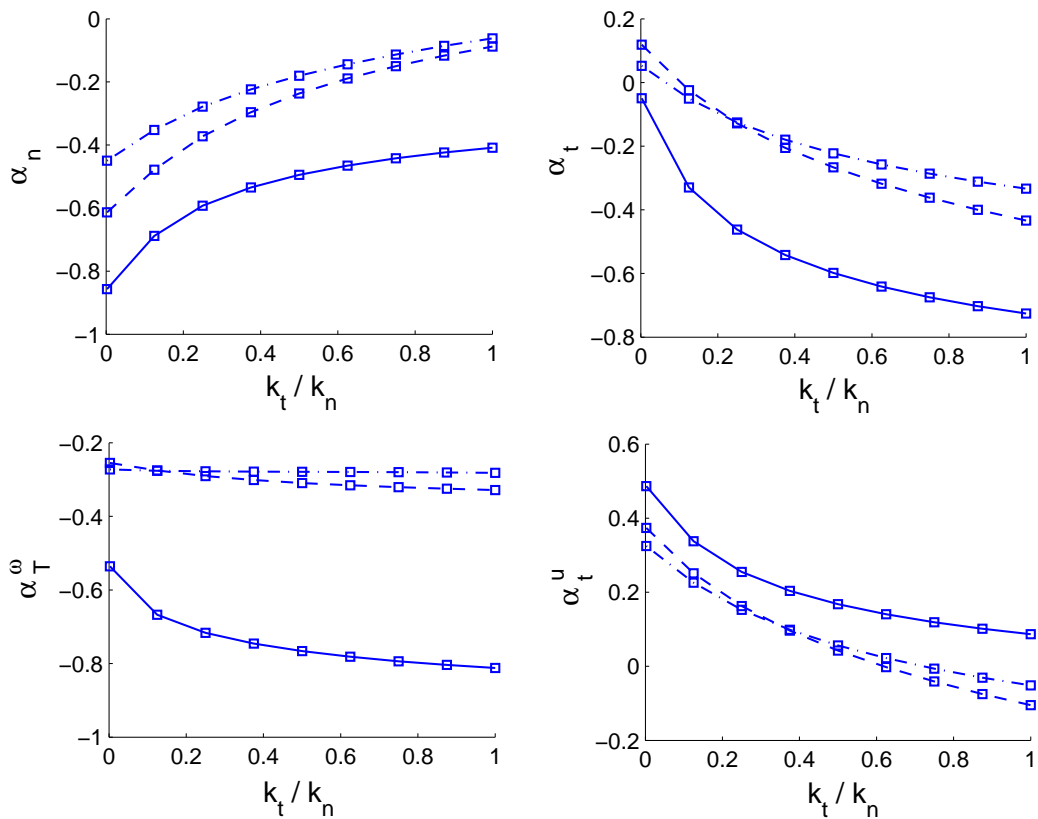}
\caption{Comparison between $\alpha _{n}$, $\alpha _{t}$, $\alpha _{t}^{u}$
and $\alpha _{t}^\omega $ from DEM simulation (solid line), 1PF (dashed) and
PF-theory (dashed-dotted). Results for $\Gamma =3.5$ and various stiffness
rations $k_{t}/k_{n}$.}
\label{alphaTheory}
\end{figure}
The predictions that include the fluctuations result in group-averages
analogous to those of Figure \ref{alphaDEM}. The
corresponding factors $\alpha _{n}$, $\alpha _{t}$, $\alpha _{t}^{u}$ and $%
\alpha _{t}^{\omega }$ are plotted in Figure \ref{alphaTheory} for
coordination number $\Gamma =3.5$. Their trend stays unchanged for
the other coordination numbers considered. The estimates
qualitatively reproduce the observed dependence of the fluctuations
on $k_{t}/k_{n}$ and on coordination number $\Gamma $. However, the
relaxation they induce is underestimated, and the stiffening
observed in the case of the particle displacement generally
overestimated. The difference between estimate and measured values
at low coordination number in the case of shear emphasizes the
necessity of considering larger correlation lengths if reliable
predictions of the mechanical behavior of such systems are to be
obtained. Given the remarkable improvement already obtained with
respect to the average strain assumption, it might be sufficient to
incorporate the fluctuations of the first neighbors.

\section{Discussion and perspectives}

This study has emphasized the insufficiency of the average strain
assumption in predicting the elastic moduli of granular media,
especially in presence of shear, as particles undergo important
additional displacements and rotations in order to attain
equilibrium. This particularly holds in the case of coordination
numbers smaller than that corresponding to the onset of
iso-staticity for frictionless assemblies, whose occurrence seems
relevant to practical purposes.

We have shown that the role of the fluctuations is clearly phrased in terms
of the relative displacements they induce at the contact points, coherently
with the constitutive law for the contact forces. The relative displacements
due to the fluctuations are highly correlated with contact orientation.
Along the normal and the tangential direction, respectively, their average
over equally oriented contacts is proportional to the relative displacements
aligned with them and due to the average strain. Such a proportionality
is expressed by the numerical factors $\beta_{n}$ for the case of isotropic compression, and $\alpha _{n}$, $\alpha _{t}^{u}$ and $%
\alpha _{t}^{\omega }$ in the case of shear. Their numerical value
allows one to interpret the role of the different kinematic
ingredients to the relaxation observed at the macroscopic scale. We
have found that the normal component of the center displacements and
the rotations always oppose the effect of the average strain. On the
contrary, the tangential component of the center displacements
generally stiffens the assembly, with the exception of large ratios
$k_{t}/k_{n}$ at large coordination number.

We have analyzed two approaches that include the fluctuations from
the average strain into the prediction of the elastic moduli, namely
the ''particle-fluctuation'' and the ''pair-fluctuation'' method.
They both determine the fluctuations by considering the problem of
equilibrium of a small subassembly. That is, they are based on the
assumption that the fluctuations organize with short correlation
length, in the order of three or four diameters. The comparison with
the DEM simulations shows a significant improvement in predicting
the moduli with respect to the average strain assumption. The
deformation mechanisms are qualitatively captured, even at small
coordination numbers, together with their dependence on the ratio
between tangential and normal stiffness and on coordination number,
thus proving the correctness of the approach. However, the models do
not capture with sufficient accuracy the fluctuations at low
coordination numbers when a shear loading is applied (especially for
the tangential relative displacements). This may be caused by the
occurrence of larger correlation lengths than those assumed here.

Future work will focus on identifying the correct correlation
length. An other open issue is the mechanical behavior at ratios of
tangential to normal stiffness larger than one, that easily occur at
the contact between
cylinders far from the onset of sliding. The independence of the $%
\alpha $'s on the contact orientation assesses that the fluctuations
that determine the macroscopic behavior originate in the average
geometry of the assembly. Therefore, analytical approaches can
capture them. As the average geometry is disordered, its
representation has to be resolved in statistical terms, as done in
the analytical solution in \cite{Agnolinetal2005}. Even though the
issue has not been dealt with in this manuscript, we anticipate that
the cited analytical prediction gives less accurate results than the
numerical implementation of the corresponding approach. This
emphasizes the sensitivity of the modeling to the details of the
statistical representation. This representation is still a critical
issue whose improvement we will pursue.

\section{Acknowledgments}

The authors acknowledge financial support of the first author, during her
stay at the University of Twente, by IMPACT, the research institute of the
University of Twente on Mechanics, Processes and Control. Financial support
has also been given by "Gruppo Nazionale di Fisica Matematica" of the
"Istituto Nazionale di Alta Matematica" through the Research Project
"Constitutive Models for Granular Materials".


\begin{thebibliography}{99}
\bibitem{Poritsky}  Poritsky, H. (1950). Stresses and deflections of
cylindrical bodies in contact with application to contact of gears and of
locomotive wheels, Journal of Applied Mechanics 17: 191-201.

\bibitem{Mindlin2}  Mindlin, R.D., Deresiewicz, H. (1953). Elastic spheres
in contact under varying oblique forces, Journal of Applied Mechanics 20:
327-344.

\bibitem{Johnson}  Johnson, K.L. (1985). Contact Mechanics. Cambridge
University Press, Cambridge.

\bibitem{Rothenburg1980}  Rothenburg, L. 1980. Micromechanics of idealised
granular materials. PhD Thesis Department of Civil Engineering, Carleton
University, Ottawa, Ontario, Canada.

\bibitem{Digby}  Digby, P.J. (1981). The effective moduli of porous granular
rock. Journal of Applied Mechanics 48: 803-808.

\bibitem{Christoffersen}  Christoffersen, J., Mehrabadi, M.M., Nemat-Nasser, S. (1981). A micro-mechanical description of granular material
behaviour, Journal of Applied Mechanics 48: 339-344.

\bibitem{Walton}  Walton, K. (1987). The effective elastic moduli of a
random packing of spheres, Journal of the Mechanics and Physics of Solids
35: 213-226.

\bibitem{BathurstRothenburg1988a}  Bathurst, R.J., Rothenburg, L. (1988a).
Micromechanical aspects of isotropic granular assemblies with linear contact
interactions. Journal of Applied Mechanics (Transactions of the ASME) 55:
17-23.

\bibitem{BathurstRothenburg1988b}  Bathurst, R.J., Rothenburg, L. (1988b).
Note on a random isotropic granular material with negative Poisson's ratio.
International Journal of Engineering Science 26: 373-383.

\bibitem{Changetal1990}  Chang, C.S., Misra, A., Sundaram, S.S. (1990).
Micro-mechanical modelling of cemented sands under low amplitude
oscillations. G\'{e}otechnique 40: 251-263.

\bibitem{Changetal1995}  Chang, C.S., Chao, S.J., Chang, Y. (1995).
Estimates of elastic moduli for granular material with anisotropic random
packing structure. International Journal of Solids and Structures 32:
1989-2008.

\bibitem{Cambouetal}  Cambou, B., Dubujet, P., Emeriault, F., Sidoroff, F. (1995). Homogenisation for granular materials. European Journal
of Mechanics A / Solids 14: 255-276.

\bibitem{Makseetal}  Makse, H.A., Gland, N., Johnson, D.L., Schwartz,
L.M. (1999). Why effective medium theory fails in granular materials.
Physical Review Letters 83: 5070-5073.

\bibitem{RothenburgKruyt2001}  Rothenburg, L., Kruyt, N.P. (2001). On
limitations of the uniform strain assumption in micromechanics of granular
materials. Powders and Grains 2001, pp.191-194, ed. Y. Kishino, Balkema
Publishers, Rotterdam, The Netherlands.

\bibitem{Suiker}  Suiker, A.S.J., Fleck, N.A. (2004). Frictional collapse
of granular materials. Journal of Applied Mechanics 71: 350-358.

\bibitem{Jks05}  Jenkins, J.T., Johnson, D., La Ragione, L., Makse, H.
(2005). Fluctuations and the effective moduli of an isotropic, random
aggregate of identical, frictionless spheres. Journal of the Mechanics and
Physics of Solids 53: 197-225

\bibitem{KruytRothenburg2004}  Kruyt, N.P., Rothenburg, L. (2004).
Kinematic and static assumptions for homogenization in micromechanics of
granular materials. Mechanics of Materials 36: 1157-1173.

\bibitem{Agnolinetal2005}  Agnolin, I., Jenkins, J.T., La Ragione L.
(2005). A continuum theory for a random array of identical, elastic,
frictional disks. Mechanics of Materials, accepted for publication
(available on line).

\bibitem{Bideau}  Gervois, A., Bideau D. (1992).
Some geometrical properties of two-dimensional hard disks packings. Disorder and granular media, Ed. Bideau and Hansen, North-Holland, 1-31.

\bibitem{Koenders}  Gaspars, N., Koenders, M.A. (2001). Micromechanical
formulation of macroscopic structures in a granular medium. Journal of
Engineering Mechanics 127: 987-992.

\bibitem{Cundall}  Cundall, P.A., Strack, O.D.L. (1979). A discrete
numerical model for granular assemblies. G\'{e}otechnique 9: 47-65.

\bibitem{CundallJenkins}  Cundall, P.A., Jenkins, J.T., Ishibashi, I.
(1989). Evolution of elastic moduli in a deforming granular assembly.
Powders and Grains 1989, Ed. J. Biarez, R. Gouvres.

\bibitem{AgnPrep}  Agnolin, I., Roux, J.N. Elastic moduli of numerical
assemblies of spheres: the role of the fluctuations from the average strain.
In preparation.

\bibitem{Drescher}  Drescher, A., de Josselin de Jong, G. (1972).
Photoelastic verification of a mechanical model for the flow of a granular
material. Journal of the Mechanics and Physics of Solids 20 337-351.

\bibitem{Strack}  Strack, O.D.L., Cundall, P.A. (1978). The distinct
element method as a tool for research in granular media: part I. Report
National Science Foundation, NSF Grant ENG75-20711.

\bibitem{RothenburgSelvadurai}  Rothenburg, L., Selvadurai, A.P.S. (1981).
A micromechanical definition of the Cauchy stress for particulate media. In:
Proceedings International Symposium on Mechanical Behaviour of Structured
Media, pp. 469-486, ed. A.P.S. Selvadurai, Ottawa, Canada.

\bibitem{Love}  Love, A.E.H. (1944). A treatise on the mathematical theory
of elasticity. Dover Publications, New York.

\bibitem{BathurstRothenburg1990}  Bathurst, R.J., Rothenburg, L. (1990).
Observations on stress-force-fabric relationships in idealized granular
materials. Mechanics of Materials 9: 65-80.

\bibitem{Calvetti}  Calvetti, F., Emeriault, F. (1999). Interparticle force
distribution in granular materials: link with the macroscopic behavior.
Mechanics of Cohesive-Frictional Materials 4: 247-279.

\bibitem{RothenburgBathurst1989}  Rothenburg, L., Bathurst, R.J. (1989).
Analytical study of induced anisotropy in idealized granular materials.
Geotechnique 39: 601-614.

\bibitem{Horne}  Horne, M.R. (1965). The behaviour of an assembly of
rotound, rigid, cohesionless particles I and II. Proceedings of the Royal
Society London A 286: 62-97.

\bibitem{KruytRothenburg2001}  Kruyt, N.P., Rothenburg, L. (2001).
Statistics of the elastic behaviour of granular materials. International
Journal of Solids and Structures 38: 4879-4899.

\bibitem{Jenkins99}  Jenkins, J.T., La Ragione, L. (1999). Particle spin in
anisotropic granular materials. International Journal of Solids and
Structures 38: 1063-1069.

\bibitem{KruytRothenburg2002}  Kruyt, N.P., Rothenburg, L. (2002).
Micromechanical bounds for the elastic moduli of granular materials.
International Journal of Solids and Structures 39: 311-324.

\bibitem{AgnolinRoux}  Agnolin, I., Roux, J.N. (2005). Elasticity of
sphere packings: pressure and initial state dependence. Powders \& Grains
2005, pp.87-91, eds. R. Garcia-Rojo, H.J. Hermann, S. McNamara, Balkema
Publishers, Rotterdam, The Netherlands. \newline
\end{thebibliography}
\end{document}